# Outcome Modeling Using Clinical DVH Data

J.J. Gordon

Department of Radiation Oncology, Henry Ford Health System, Detroit MI, U.S.A.

**Purpose:** To quantify the ability of correlation and regression analysis to extract the normal lung dose-response function from dose volume histogram (DVH) and radiation pneumonitis (RP) data.

**Methods:** A local injury model is adopted, in which radiation-induced damage (functional loss) G is the integral of the DVH with function R(D). RP risk is H(G) where H() is the sigmoid cumulative distribution of functional reserve. RP incidence is a Bernoulli function of risk. A homogeneous patient cohort is assumed, allowing non-dose-related factors to be ignored. Clinically realistic DVHs are combined with the injury model to simulate RP data.

**Results:** Correlation analysis is often used to identify a subset of predictor variables that are significantly correlated with outcome, for inclusion in a predictive model. In the local injury model, all DVH metrics VD contribute to damage through the integral with R(D). Correlation analysis therefore has limited value. The subset of VD that are most significantly correlated with incidence varies randomly from trial to trial as a result of random variations in the DVH set, and does not necessarily reveal anything useful about the patient cohort or the underlying biological dose-response relationship. Regression or matrix analysis has the potential to extract R(D) from damage or risk data, provided smoothness regularization is employed. Extraction of R(D) from incidence data was not successful, due to its higher level of statistical variability.

**Conclusions:** To the authors' knowledge, smoothness regularization has not been applied to this problem, so represents a novel approach. Dose-response functions can be successfully extracted from measurements of integral (as opposed to regional) lung damage G, suggesting value in re-visiting available measurements of ventilation, perfusion and radiographic damage. The techniques developed here can potentially be used to extract the dose-response functions of different tissues from multiple types of quantitative volumetric imaging data.

*This manuscript will be submitted to a journal.*

## I. INTRODUCTION

Forward progress in radiation therapy (RT) depends on the ability to establish rigorous causal relationships between patient characteristics and treatment parameters on the one hand, and treatment outcomes on the other. Patient characteristics include comorbidities, performance status, genetic markers, etc. Treatment parameters include first and foremost the planned 3D dose distribution, but also fractionation, concurrent chemotherapy, immobilization and motion management. Outcomes include overall survival, local tumor control, and a variety of site-specific toxicities and complications that potentially degrade patients' quality of life.

A key pitfall is to attribute clinical significance to results that are in fact incidental artifacts of either the specific dataset, or the analysis technique. This work uses radiation pneumonitis (RP), one of the principal complications of lung cancer RT, to illustrate the challenges of extracting dose-response relationships from clinical dose-volume histogram (DVH) data.[1] A local damage / injury model similar to Jackson et al[2] is adopted. It is assumed that normal lung is a parallel organ, and that radiation-induced loss of function can be modeled using a biological dose-response profile P(D), which gives the probability of sub-volume damage as a function of local dose D.

Integral damage G is the integral of the differential DVH with P(D), or of the cumulative DVH with $R(D) \equiv P'(D)$. RP risk (or jeopardy) J is assumed to be a sigmoid function of G, modeled for example using the Lyman-Kutcher-Burman equations.[3] RP incidence I is a binary variable indicating whether RP above some threshold was (was not) experienced. I is distributed according to the Bernoulli distribution B(J) (i.e., binomial distribution for a single trial having probability of success J). For the purposes of this study, non-dose-related factors are ignored — damage G, and therefore risk J and incidence I, are assumed to depend solely on the DVH. For the extraction of dose-response relationships, this represents a best-case scenario. Clinical RP data is likely to reflect a range of non-dose factors, and will therefore be more challenging.

DVHs from the University of Michigan (UM) dose escalation trial reported by Kong[4] are used to generate multiple random sets of clinically realistic lung DVHs, simulating the results of multiple clinical trials. The DVHs are combined with two hypothetical dose-response profiles P(D) to obtain simulated RP data. Correlation and regression analysis is then applied to the simulated RP data to see how well different analytical techniques can recover the known dose-response profile. Specifically, damage is approximated as a weighted sum over dose metrics $VD_i$, where $VD_i$ is the volume of normal lung receiving dose of $D_i$ or greater, calculated at doses $D_i = 5, 10, 15, ... , 100$ Gy. Statistical analysis attempts to recover the weights multiplying $VD_i$.

We find that correlation analysis provides little useful information regarding the underlying dose-response. The set of $VD_i$ that are most significantly correlated with RP varies randomly between trials, and is unrelated to the contribution of $VD_i$ to damage G. No significance can be read into the fact that, e.g., V5 or V13 is significantly correlated with RP in some trials, but not in others. Regression analysis is able to extract P(D) from damage or risk data, but not from incidence data. The statistical variability inherent in incidence data challenges the regression algorithms. In order to successfully extract P(D) from damage or risk data, appropriate regularization (smoothness penalties) must be employed. Conventional L1 and L2 regularization are ineffective.

In summary, to have a good prospect of extracting P(D) from DVH data, surrogate measurements of damage G should be incorporated into clinical trials. In the case of normal lung, candidate surrogates include perfusion and diffusion loss, and radiographic damage. This work employs a novel smoothness regularization technique that has not previously been applied to this problem. Without this technique, regression analysis returns noisy results, due to the mathematically ill-posed nature of the regression problem. Provided accurate estimates of integral damage are available, the techniques developed here can reliably recover the dose-response functions P(D) and R(D) from DVH and damage data. Extracting the dose-response function from incidence data is more challenging.

## II. MATERIALS AND METHODS

### A. University of Michigan DVH Data

UM data is reported in detail by Kong [4] (see also prior publications cited by Kong), and has been previously analyzed in Gordon et al. [5,6] In summary, the UM dataset consists of 89 DVHs for total lung minus gross tumor volume (GTV). Treatments were either RT alone or RT with neoadjuvant chemotherapy. RT planning employed 3D conformal RT (3DCRT). Patients received daily fractions of 2.1 Gy, with total dose escalated from 63 Gy based on normal tissue tolerance. Grading of RP was according to SWOG (www.swog.org) rules, with clarifications noted in Kong. Data were provided in anonymized form for patients who had known RP status at a time point 6 months after the start of treatment. RP grading was as follows (G = grade): G0: 42, G1: 30, G2: 12, G3: 5, G4+: 0. DVHs were given as absolute lung volumes (in cubic centimeters (cc)) in dose bins with centers from 0.5 Gy to 119.5 Gy in steps of 1 Gy, calculated using the UMPlan treatment planning system. DVH dose values were provided as 2 Gy equivalent (EQD2) based on an α/β ratio of 2.5 Gy. The present study converted EQD2 doses to physical doses before integrating with damage profiles.

Figure 1 shows a plot of mean lung dose (MLD) versus DVH number for the University of Michigan dataset. Actual radiation pneumonitis (RP) incidence, derived from the clinical trial, is indicated by symbol. Pink solid triangles are grade 3 DVHs. Red solid circles are grade 2 DVHs. Blue hollow diamonds are grade 1 DVHs. Gray hollow squares are grade 0 DVHs. Note that these toxicity gradings are ignored in the simulated trials, to be described below. In the simulations, damage, risk and incidence values are generated using the equations given in sections $2.2 - 2.4$. The gray band indicates DVHs having 15 Gy $\leq$ MLD $\leq$ 20 Gy. DVHs from this MLD range are used to illustrate differences in correlation behavior between DVH subsets and the full DVH set.

### B. Simulated DVHs

To simulate a clinical trial, some number N (e.g., N = 100) of simulated DVHs are generated from the UM DVHs, either from the full set, or from the subset having 15 Gy $\leq$ MLD $\leq$ 20 Gy. To generate each simulated DVH (DVHsim), two randomly selected UM DVHs (DVHUM1 and DVHUM2) are combined with a uniformly-distributed random weight:

$$DVH_{sim}(d) = \mu \cdot DVH_{UM1}(d) + (1 - \mu) \cdot DVH_{UM2}(d) \quad (1)$$

where $\mu \in [0,1]$ is the weight. By default all simulated DVHs are assumed to be associated with conventional 30-fraction treatment courses. However, results are not expected to depend on the assumed fractionation, and will apply more generally to other fractionation schemes.

### C. NTCP Model

In the following DVH(D) and cDVH(D) denote differential and cumulative DVHs: $DVH(D) = (d/dD) \, cDVH(D)$, $cDVH(D) = \int_D^{\infty} DVH(d)$. This work adopts a local damage / injury model similar to Jackson et al. [2] (See also Niemierko [7] and Rutkowska. [8]) It is assumed that lung is a parallel organ, and that radiation-induced loss of function (damage) can be modeled using a probability profile P(D), which gives the probability of sub-volume damage as a function of local dose D. Integral damage G is obtained by integrating DVH(D) · P(D). Mathematically this is equivalent to integrating cDVH(D) · R(D), where R(D) is the derivative of P(D): R(D) = (d/dD) P(D) ≡ P'(D). For this work, functions P(D) are referred to as damage profiles, and functions R(D) are referred to as rate profiles.

$$G = \int_0^{\infty} DVH(D) \cdot P(D) \; dD$$
$$= \int_0^{\infty} cDVH(D) \cdot R(D) \; dD \quad (2)$$

Radiation pneumonitis is assumed to occur when G exceeds the lungs' functional reserve. As in Jackson et al [2], RP risk or jeopardy, denoted by J, is given by J = H(G), where H(G) is the cumulative distribution of functional reserves in the patient population. Based on clinical data, H(G) is a sigmoid function, modeled e.g., using the LKB equation [3]

$$J = H(G) \approx LKB(G_{50}, m_G, G) \qquad (3)$$

where $G_{50}$ is the damage value corresponding to 50% risk of RP, and $m_G$ is the LKB slope parameter. For a specific patient, incidence I is a binary random variable taking the value 0 (RP absent) with probability $(1 - J)$, or the value 1 (RP present) with probability J

$$I \sim B(J) \qquad (4)$$

where $B(p)$ is the Bernoulli distribution (i.e., binomial distribution for a single trial having probability of success $p$). Equivalently, I is 0 if the patient's functional reserve is greater than or equal to G, 1 if it is less than G. For his work we further define "RP" to mean clinically diagnosed RP of grade 2 or higher (G2+) according to SWOG criteria. I = 0 (or 1) therefore signifies G1– (or G2+) RP. However, results are not dependent on this definition, and will apply also to alternative RP definitions.

### D. Simulated RP Data

Simulated damage values G are calculated from the DVHs in section 2.2 by combining them with hypothetical damage profiles P(D) (or R(D)). This work considers two types of damage profile: linear quadratic (LQ) [3] profiles having the form

$$P(D) = 1 - \exp(-D(\alpha + \beta D/N)) \qquad (5)$$

and Joiner's induced repair model (IRM) [9] which modifies the LQ model by adding an initial low-dose section with steeper slope, followed by a section with shallower slope, before asymptoting to LQ behavior. The original IRM equations model single-fraction cell survival curves, but may be generalized to multiple fractions by assuming that, similar to LQ curves, the N-fraction cell survival curve $S_N(D)$ is given by $S_N(D) = S_1(D/N)^N$, where $S_1(D)$ is the single-fraction survival curve. The generalized IRM equations for the damage profile $P(D) = 1 - S_N(D)$ are:

$$P(D) = 1 - \exp(-D[\alpha(D/N) + \beta D/N]) \qquad (6)$$

$$\alpha(D) = \alpha_r - (\alpha_r - \alpha_s) \cdot \exp(-D/D_c) \qquad (7)$$

where, as in Joiner's original formulation, $\alpha_s$ is the initial slope and $\alpha_r$ is the final slope of the single-fraction cell survival curve, $D_c$ is a critical dose that determines the transition from $\alpha_s$ to $\alpha_r$, and $\beta$ is the usual LQ parameter.

This work adopts the following LQ model parameters: $\alpha = 0.01$ $Gy^{-1}$, $\beta = 0.005$ $Gy^{-2}$ and $\alpha/\beta = 2$ Gy. The $\alpha$ value is derived from the in-vivo SPECT perfusion loss measurements reported by Koontz et al. [10] Recent measurements [10,11] suggest that the $\alpha/\beta$ ratio for normal lung is lower than 3 Gy, motivating the value $\alpha/\beta = 2$ Gy. Based on Gordon et al. [5] selected IRM model parameters are $\alpha_s = 0.04$, $\alpha_r = 0.0032$, $D_c = 0.22$, $\beta = 0.0032$, and $\alpha_r/\beta = 1$. The default assumption in this work is that the number of fractions N = 30. Figure 2 plots P(D) and R(D) for LQ and IRM profiles having N = 30 and the above parameter values.

Damage values are translated into risk values using the LKB model (equation (3)). For LQ profiles we assume $G_{50} = 0.27$ and $m_G = 0.25$. For IRM profiles we assume $G_{50} = 0.18$ and $m_G = 0.25$. (Note that a $G_{50}$ value of 0.27 (0.18) implies that RP risk is 50% when 27% (18%) of normal lung is damaged.) For conventionally fractionated RT, Borst et al [12] report crude incidence rates of RP to be 17.6%. The LKB parameters are selected to give RP incidence of around 17%, consistent with the clinical values reported by Borst.

The LQ model is a generally accepted model of radiation-induced cell damage. Motivation for the alternative IRM model is given by Gordon et al. [5,6] For the LKB model, parameter fitting could be done in different ways. For example, parameters could be derived from single-fraction whole-lung irradiation data. [13] However, the intention of this work is not to fit model parameters to any specific clinical dataset, but rather to use clinically realistic values to illustrate the problem of extracting dose-response curves from DVH data. The selected LQ, IRM and LKB models provide a realistic test of analysis techniques' ability to extract P(D) and R(D).

### E. Analytical Techniques

This section explains how covariance and correlation values are calculated between dose metrics VD and RP data (G, J or I). Additionally, it explains how one may solve for R(D) using covariance equations, or regression methods.

#### E.1. Covariance / Correlation Analysis

Results of clinical trials are frequently reported in the form of correlations between RP and dose metrics such as MLD, V20, V30, etc. Practical analysis typically uses a discrete set of doses, e.g., $D_j = 5, 10, ..., 100$ Gy, and corresponding metrics $VD_j = V5, V10, ... , V100$. In the following it is assumed that a trial has accrued N patients, produc-ing DVHs $DVH_n(d)$, with corresponding metrics $VD_{n,j}$, damage values $G_n$, risk values $J_n$, and incidence values $I_n$, $n = 1, ..., N$. To simplify presentation, the index $n$ is suppressed. Correlations and covariances are understood to be evaluated over the DVH set (i.e., index $n$). From equation (2), damage can be approximated as a sum

$$G \approx \sum VD_j \cdot R_j \qquad (8)$$

where the sum is over j, $R_j \equiv R(D_j) \cdot \Delta D$, and $\Delta D$ is the dose bin size, here equal to 5 Gy. Using equation (8), $cov(VD_i, G)$ can be approximated

$$cov(VD_i, G) \approx \sum cov(VD_i, VD_j) \cdot R_j \qquad (9)$$

Note that the function H() and its inverse are assumed unknown. However, if $L(p) = \ln(p/(1-p))$ is the logit function, then $L(J)$ closely approximates a linear function of G

$$L(J) \approx \kappa + \lambda \cdot G \approx \kappa + \lambda \cdot \sum VD_j \cdot R_j \qquad (10)$$

where $\kappa$ and $\lambda$ are unknown constants. Combining (9) and (10) gives

$$cov(VD_i, L(J)) \approx \lambda \cdot \sum cov(VD_i, VD_j) \cdot R_j \qquad (11)$$

This work performs all analysis within Matlab®. Correlations and covariances are computed using Matlab's $corr()$ and $cov()$ functions.

### E.2. Matrix Solutions

If $\tilde{c}$ is defined to be the column vector of covariances $cov(VD_i, G)$, $\tilde{l}$ the column vector of covariances $cov(VD_i, L(J))$, $\tilde{r}$ the column vector of $R_j$, and $M \equiv [M_{ij}]$ the covariance matrix $cov(VD_i, VD_j)$, then equations (9) and (11) may be written in matrix form

$$\tilde{c} = M\tilde{r} + \tilde{\epsilon}_c \qquad (12)$$

$$\tilde{l} = \lambda M\tilde{r} + \tilde{\epsilon}_l \qquad (13)$$

where $\tilde{\epsilon}_c$ and $\tilde{\epsilon}_l$ are error vectors that convert the approximations (9) and (11) to equalities. We focus on equation (12), but the following comments also apply with adaptation to (13). Intuitively, the elements of $\tilde{\epsilon}_C$ can be made arbitrarily small by using many closely-spaced dose metrics $VD_j$ in (8). If $\tilde{\epsilon}_C$ is zero, equation (12) is easily solved: $\tilde{r} = M^{-1}\tilde{c}$. If $\tilde{\epsilon}_C$ is small but non-zero, one can formally solve the matrix equation, but a problem arises. If the matrix

M is close to being degenerate, the problem is referred to as ill-posed. In that scenario, small amounts of noise cause the formal solution of (12) to diverge widely from the true (zero-noise) solution. Additionally, small differences in the noise cause large variations in the formal solution, making the formal solution useless.

As will be demonstrated below, the present problem — extracting P(D) and R(D) from DVH data — is ill-posed. Many real-world problems are ill-posed, requiring regularization. Regularization refers to the strategy of adding a penalty term to the problem, which has the effect of reducing its sensitivity to noise, allowing the solution of the modified (regularized) problem to closely approximate the solution of the original zero-noise problem. Here we utilize Tikhonov regularization, which solves the matrix problem (12) by minimizing $\|M\tilde{r} - \tilde{c}\|^2 + \|\Gamma\tilde{r}\|^2$, where $\Gamma$ is a Tikhonov (penalty) matrix. A common choice is $\Gamma = \alpha I$, where $\alpha$ is a small scalar and I is the identity matrix. This choice penalizes the vector norm of $\tilde{r}$, selecting a solution that has fewer non-zero elements. A more appropriate choice for the present problem is

$$\Gamma = \frac{\alpha}{2} \begin{bmatrix} 1 & -1 & & 0 \\ & 1 & -1 & \\ & & \ddots & \ddots \\ 0 & & 1 & -1 \end{bmatrix} \qquad (14)$$

As described in Reichel and Ye [14], this matrix is a finite difference approximation to a derivative. It penalizes jaggedness in $\tilde{r}$, selecting a smooth solution. The Tikhonov regularized solutions of equations (12) and (13) are [15]

$$\tilde{r} = (M^T M + \Gamma^T \Gamma)^{-1} M^T \tilde{c} \qquad (15)$$

$$\tilde{r} = \lambda^{-1} (M^T M + \Gamma^T \Gamma)^{-1} M^T \tilde{l} \qquad (16)$$

Equation (15) determines $\tilde{r}$ exactly. Equation (16) determines $\tilde{r}$ up to an unknown multiplier $\lambda$.

### E.3. Regression Solutions

A generalized linear model (GLM) [16] expresses a dependent variable $y$ in terms of a linear combination of $m$ independent variables $x$, via a link function $f()$

$$Y = f^{-1}(Xw) + \epsilon \qquad (17)$$

In equation (17), Y is an n x 1 column vector of observations of $y$, X is an n x m matrix of observations of $x$, $w$ is a column vector of weights and $\epsilon$ represents noise. The problem is to extract the weights $w$ given observations Y and X containing some degree of noise. Equation (8) is in the form (17), with $y \leftrightarrow G$, $x \leftrightarrow VD_j$, $w \leftrightarrow R_j$ and identity link function. Equation (10) is also in the form (17), with $y \leftrightarrow L(J)$, $x \leftrightarrow VD_j$, $w \leftrightarrow R_j$ and identity link function. Finally, RP incidence I may be expressed in the form (17), where $y \leftrightarrow I$, $x \leftrightarrow VD_j$, $w \leftrightarrow R_j$ and $f()$ is the logit or inverse LKB function.

All of these problems may be solved using regression algorithms designed for GLM problems. This work reports on the use of Matlab's *glmfit()* and *lassoglm()* functions. It also uses Mineault's *glmfitqp()* function. [17] The Matlab functions attempt to extract best-fit weights $w$ by minimizing the norm of $\epsilon$. The *glmfitqp()* function uses a maximum likelihood approach, and allows for a quadratic smoothness penalty $\|\Gamma w\|^2$ with $\Gamma$ as in equation (14).

## III. RESULTS

### A. Covariance / Correlation Analysis

Figure 3 plots $cov(VD_i, VD_j)$ and $corr(VD_i, VD_j)$ to illustrate the strong correlations that exist between dose metrics VD within a clinical DVH dataset. Fig. 3a-b plots $cov(V30, VD_j)$ and $corr(V30, VD_j)$ for $D_j = 5, 10, \ldots, 100$ Gy. The gray lines represent the covariance and correlation (CC) for 10 simulated clinical trials, each consisting of 100 DVHs. The central solid blue line is the average. The outer dashed blue lines represent the average ± one standard deviation. Fig. 3c-d plots mean $cov(VD_i, VD_j)$ and mean $corr(VD_i, VD_j)$ (10 trials x 100 DVHs) for the case where simulated DVHs are generated from the full UM DVH set. Fig. 3e-f plots mean $cov(VD_i, VD_j)$ and mean $corr(VD_i, VD_j)$ (10 trials x 100 DVHs) for the case where simulated DVHs are restricted to the MLD range 15 Gy ≤ MLD ≤ 20 Gy.

Figure 4 plots $cov(VD, G)$, $corr(VD, G)$, $cov(VD, I)$ and $corr(VD, I)$. Plots of $cov(VD, J)$ and $corr(VD, J)$ are visually similar to plots of $cov(VD, G)$ and $corr(VD, G)$, so are not shown. Fig. 4a-d are for the case where simulated DVHs are generated from the full UM DVH set. Fig. 4e-g are for the case where simulated DVHs are restricted to the MLD range 15 Gy ≤ MLD ≤ 20 Gy. All results were obtained by simulating 10 trials x 100 DVHs. The gray lines show the results for individual trials. The solid gray and blue lines are for the LQ dose-response model. The dotted gray and red solid lines are for the IRM dose-response model.

Figure 5 plots the percentage of trials in which VD has a statistically significant ($p < 0.05$) correlation with G or I, and the percentage of trials in which VD has the most significant (smallest p-value) correlation with G or I. These results were obtained by simulating 100 trials x 100 DVHs. Note that plots of correlations with J are visually similar to those for G, so are not shown. Fig. 5a-d are for the case where simulated DVHs are generated from the full UM DVH set. Fig. 5e-g are for the case where simulated DVHs are restricted to the MLD range 15 Gy ≤ MLD ≤ 20 Gy. Blue bars are for the LQ dose-response model. Red bars are for the IRM dose-response model.

### B. Matrix Solutions

Figs. 6a-b show solutions of the matrix equation (12) for the case where there is zero noise. To generate these results, zero-noise profiles $R_{ZN}(D)$ were defined as follows:

$$R_{ZN}(D) = \sum_{d=5}^{80} R(d)\, \delta(d - D) \qquad (18)$$

$R(D)$ is the LQ or IRM profile from Fig. 2b. The LQ profile is reproduced as the blue line in Fig. 6a. The IRM profile is reproduced as the red line in Fig. 6b. The sum in (18) is over the discrete set of doses d = 5, 10, ..., 80 Gy. The delta functions ensure that $R_{ZN}(D)$ is non-zero at only those doses, causing the approximation in equation (8) to be exact, and the noise $\bar{\epsilon}_c$ in equation (12) to be zero. Fig. 6 is based on 10 trials x 100 DVHs. The gray lines in Figs. 6a-b are the solved profiles: $\bar{r} = M^{-1}\,\bar{c}$. For all trials, the solution exactly matches (18) — gray lines for the 10 trials overlie one another.

Figs. 6c-d show solutions $\bar{r} = M^{-1}\,\bar{c}$ for the case where profiles $R_{ZN}(D)$ are still employed, but where 1% uniformly distributed noise is artificially added to G. Because the covariance matrix M is close to being degenerate, the problem is ill-posed and that small amount of noise is enough to make the solved profiles $\bar{r}$ incorrect. If no noise is added to G, but the true profiles R(D) are employed in place of the zero-noise profiles $R_{ZN}(D)$, one obtains plots that are very similar to Figs. 6c-d. The approximation of the continuous integral (2) by the discrete sum in (8) introduces enough noise to produce erroneous estimates of R(D).

Figs. 6e-f show the Tikhonov regularized solutions (15) for the case where the true profiles $R(D)$ are used to calculate G, no additional noise is added, dose metrics V5, V10, ..., V100 are included in the covariance matrix, and the Tikhonov multiplier $\alpha$ in equation (14) is equal to 0.001. The regularized solution does an adequate job of reproducing the true profiles up to 100 Gy.

The results in Fig. 6 were obtained by simulating 10 trials x 100 DVHs, using the full DVH dataset, and covariances $cov(VD, G)$ (equation(15)). If one uses the restricted DVH set with 15 Gy ≤ MLD ≤ 20 Gy, the non-regularized solutions in Figs. 6c-d are a little less variable, and the regularized solutions in Figs. 6e-f are clustered a little more tightly around the true profile, but results are otherwise similar. If one uses covariances $cov(VD, L(J))$ and solves equation (16), results are very similar — solutions using logit(J) are essentially equivalent.

### C. Regression Solutions

If regression solutions are generated for G or J vs VD for the zero-noise problem (i.e., $R(D)$ replaced by $R_{ZN}(D)$), the obtained solutions are exact, resembling Figs. 6a-b. In this case, Matlab's *glmfit()* function is able to find the solutions. If the true profiles R(D) are employed, introducing noise through the approximation (8), *glmfit()* solutions are unstable, as shown in Figs. 7a-b. If Lasso or ElasticNet regularization is attempted using Matlab's *lassoglm()* function, the algorithm fails to converge to a solution, exiting after reaching its iteration limit. If Mineault's *glmfitqp()* function is employed, stable solutions are obtained, as shown in Figs. 7c-d.

If Mineault's algorithm is used to regress I against VD, simulation of 10 trials x 100 DVHs gives the solutions in Figs. 7e-f. The extra variability in I (as compared with G or J) challenges the algorithm. If Mineault's algorithm is again used to regress I against VD, simulation of 10 trials x 1000 DVHs gives the solutions in Figs. 7g-h. Expanding the size of the trials from 100 to 1000 DVHs (i.e., patients) gives the algorithm more data to work with, ensuring somewhat more accurate results. In the case of regression, R(D) estimates obtained using a restricted subset of DVHs have noticeably more variability than solutions obtained using the full DVH set. In the case of the simulations, restricting the regression analysis to a subset of DVHs appears to be counterproductive.

### D. Regression Solution of the Original UM Data

The clinical trial data provided by the University of Michigan included DVHs and RP incidence (Fig. 1), but no surrogate measurements of damage G, and insufficient data to generate reliable estimates of risk J. Fig. 8 shows the estimates of R(D) extracted from the UM clinical trial data. These estimates were obtained by performing regression of I against VD, using the UM incidence data in Fig. 1 (G2+ vs G1– cases) and Mineault's *glmfitqp()* function.

## IV. DISCUSSION

### A. Correlation Analysis

Fig. 3d shows that when all DVHs are included in the analysis, strong correlations (i.e., correlations $\geq \sim 0.8$) exist between all dose metrics V5 – V50. In equations (11) and (12) this ensures that many $VD_i$ / $R_j$ combinations contribute strongly to G, making it more challenging to identify the impact of any single $R_j$, or to detect the differences in G resulting from different R(D) profiles. Fig. 3f shows that when analysis is restricted to DVHs matching a narrow MLD range, correlations $corr(VD_i, VD_j)$ are more narrowly peaked around $i = j$. This increases the probability of detecting the differences in G resulting from different R(D) profiles. Clinical DVH datasets tend to include large numbers of cases that have negligible RP risk. For example, there are no cases of RP for DVHs having MLD < 15 Gy in Fig. 1. Paradoxically, including large numbers of these "uninformative" DVHs in correlation analysis tends to blur correlation profiles, making it more difficult to establish a connection with the underlying dose-response profile.

This is confirmed by Fig. 4. In Figs. 4a-d, where results have been generated using the full DVH set, the LQ and IRM models exhibit similar profiles, making it difficult to discriminate the dose-response model based on the correlation profiles. In contrast, in Figs. 4e-f, where results have been generated using the restricted DVH set, the LQ and IRM models exhibit differently-shaped profiles, making it feasible to test the hypothesis that the true dose-response is e.g., LQ and not IRM. Non-parametric tests such as Kolmogorov-Smirnov could possibly be adapted for this purpose. However, discrimination is only possible using the covariance or correlation of G (or J) with VD. Correlations of I with VD (Fig. 4h) are very small, and dominated by inter-trial noise, rendering hypothesis testing impossible.

When testing for statistically significant correlations (p < 0.05) of VD with G, J or I, one is testing for non-zero correlations, without any regard to the size of the correlations. If one accepts a local injury model (equation (2)), all dose metrics VD contribute to G at some level. Therefore, it would not be surprising to find that a wide range of VD exhibit non-zero correlation. Guckenberger, for example, reports significant correlations for V2.5 – V50. [18] In the case of incidence data, Fig. 5c shows that most VD are significantly correlated with I in fewer than 100% of trials. This implies that there is some trial-to-trial variability in the set of apparently correlated VD. This is reinforced by Fig. 5d, which shows that any VD can register the most significant correlation in some subset of trials. It follows that no significance can necessarily be read into the fact that certain VD, e.g., V5 or V13, are significantly correlated with RP in some trials, but not in others. This occurs simply as a result of variations in the DVH set, and does not necessarily reveal anything useful about the patient cohort or the underlying biological dose-response relationship.

## B. Matrix & Regression Solutions

Figures 6 and 7 show the value of using a smoothness penalty when estimating R(D), rather than alternative types of regularization (Ridge, Lasso, ElasticNet) that penalize the number of non-zero elements in $\tilde{r}$, or the L1 or L2 magnitude of $\tilde{r}$. Alternative regularization methods apply the wrong type of penalty for this problem, leading to invalid results.

This work considers the ideal situation where RP is determined solely by the DVH and an (unspecified) sigmoid risk function. Real-world data will reflect other non-dose-related factors, such as the effect of comorbidities, performance status, genetic markers, chemotherapy, etc. Extraction of dose-response from real-world data may therefore be more challenging. This work provides optimism that dose-response functions can be extracted from clinical damage or risk data (G or J). However, incidence data (I) still appears too challenging,, unless trials enroll substantially larger numbers of patients than is presently the case, or DVH plus complication data from multiple trials is pooled.

## C. Dose-Response Determination

The general problem addressed in this work is how to determine the full DVH dependence of radiation-induced tissue injury, from clinical DVHs plus measures of tissue damage (G, J, or I). At the present time, data regarding tissue damage (i.e., RT complications) is most commonly supplied in the form of grading data. In the case of radiation pneumonitis, a physician classifies each patient as grade 0, 1, 2, ... based on symptoms and imaging. Grading is performed according to SWOG, CTCAE or other scheme. [19] Grading data is frequently reduced to binary incidence I by setting a severity threshold — in this work we consider G2+ versus G1– RP.

Even with strict controls, physician grading of RT complications is inevitably subjective. For example, the challenges associated with the grading of radiation pneumonitis are discussed in the Quantec paper. [1] The results of this work effectively ignore subjectivity in RP grading — simulated incidence data is generated from the NTCP model, assuming the underlying damage and risk values can be calculated exactly. The results given here show that, even without the subjectivity associated with different physicians, different institutions, uneven follow-up, and uneven imaging capabilities, the variability inherent in incidence (or grading) data makes it extremely challenging to extract DVH dependence. As long as clinical trials continue to provide results in the form of simple ordinal grading data, it seems doubtful that further quantitative insight will be provided into radiation dose-response.

This work shows that, if quantitative volumetric measures of tissue damage are available (G or J), it is feasible to determine DVH dependence using available algorithms. Fortunately, quantitative volumetric measures of tissue damage are becoming technically feasible, courtesy of functional magnetic resonance imaging (fMRI) , MR spectroscopy (MRS), diffusion weighted imaging (DWI), diffusion tensor imaging (DTI), arterial spin labelling (ASL), and other advanced imaging techniques. [20] The clinical utility of these techniques remains to be determined. However, each produces one or more candidate surrogate measurement of some aspect of tissue damage, to which the techniques of this work can potentially be applied.

The techniques developed here utilize integral, as opposed to regional, measures of tissue injury. It is reasonable to ask whether one could bypass integral measures, and use the above imaging techniques to directly measure regional tissue injury. In that case, imaging would directly measure the dose response function P(D). As imaging techniques continue to advance, this may become possible. However, direct measurement of regional damage can rely on deformable image registration (DIR), which is subject to its own uncertainties. A number of studies have attempted regional measurements, and found them to be challenging. [21,22] For lung, Liao notes that "measurements of regional lung function are difficult and pulmonary function tests most often assess total lung function". [23]

Additionally, it remains conceivable that the injury caused by radiation dose deposition in one voxel may actually manifest at a different location in the organ. In the case of lung, out-of-field injury has been clinically observed. [24,25] The mechanisms of in vivo tissue damage involve organ-level and systemic immune response, and could involve bystander signaling. [26,27] Until the various mechanisms involved in radiation induced injury are better understood, these factors may complicate accurate regional measurements. Integral injury measurements are likely to be robust to the uncertainties associated with regional measurements, and therefore could provide a more reliable method of determining DVH dependence.

The principal innovations of this work are to show that the problem of determining DVH dependence from DVHs plus injury data is ill-posed, and that smoothness regularization can counter this problem, enabling reliable solution. The ill-posedness is due to the strong correlations between dose metrics VD within clinical DVH sets. This work demonstrates two forms of smoothness regularization — Tikhonov regularization of a matrix solution, and a smoothness penalty term added to a regression solution. However, other forms are possible. For example, functional principal components analysis (FPCA) can achieve the same end, by expressing solutions in terms of smooth orthonormal basis functions. [28] The factor enabling DVH dependence to be successfully extracted is that some form of smoothness regularization be employed.

## V. CONCLUSIONS

This work is predicated on a local injury model of normal lung, in which radiation-induced damage is the integral of the cumulative DVH with a dose-response function R(D), RP risk is a sigmoid function of integral damage, and incidence is related to risk via the Bernoulli distribution. A homogeneous patient cohort is assumed, allowing all non-dose-related determinants of RP to be ignored. Using simulated DVH plus RP data, the ability of correlation and regression analysis to extract the dose-response function is examined.

In problems of this type, correlation analysis is often used to identify the subset of candidate predictor variables that are significantly correlated with the outcome, and should therefore be included in a predictive model. This approach makes little sense for the case where lung damage depends on all variables VD, courtesy of the integral with R(D). Correlation analysis therefore has limited value. It is demonstrated that the subset of VD that are significantly correlated with RP varies randomly from trial to trial. This suggests that no significance can necessarily be attached to the fact that certain VD, e.g., V5 or V13, are significantly correlated with RP in some trials, but not in others. This occurs simply as a result of random variations in the DVH set, and does not necessarily reveal anything useful about the patient cohort or the underlying biological dose-response relationship.

This work was not successful in extracting R(D) from incidence data (I), due to its higher level of statistical variability. However, regression or matrix analysis has the potential to extract R(D) from damage or risk data, provided appropriate smoothness regularization is employed. To the authors' knowledge, smoothness regularization has not previously been applied to this problem, so represents a novel approach. In particular, this work shows that dose-response functions can be successfully extracted from measurements of integral (as opposed to regional) lung damage G, suggesting value in re-visiting available measurements of ventilation, perfusion and radiographic damage. The techniques developed here can potentially be used to extract the dose-response functions of different tissues from multiple types of quantitative volumetric imaging data.

## ACKNOWLEDGEMENTS

The authors are grateful to the University of Michigan Department of Radiation Oncology for providing access to the clinical trial data used for this work.

## REFERENCES

1. Marks, L. B. *et al.* Radiation Dose-Volume Effects in the Lung. *Int. J. Radiat. Oncol. Biol. Phys.* **76,** S70–S76 (2010).
2. Jackson, A. *et al.* Analysis of clinical complication data for radiation hepatitis using a parallel architecture model. *Int. J. Radiat. Oncol. Biol. Phys.* **31,** 883–891 (1995).
3. Li, X. A. *et al.* AAPM Task Group 166 Report: The Use and QA of Biologically Related Models for Treatment Planning. *Med. Phys.* **39,** 1386–409 (2012).
4. Kong, F.-M. *et al.* Final toxicity results of a radiation-dose escalation study in patients with non-small-cell lung cancer (NSCLC): predictors for radiation pneumonitis and fibrosis. *Int. J. Radiat. Oncol. Biol. Phys.* **65,** 1075–1086 (2006).
5. Gordon, J. J. *et al.* Extracting the Normal Lung Dose-Response Curve from Clinical DVH Data: A Possible Role for Low Dose Hyper-Radiosensitivity, Increased Radioresistance. *TBD*
6. Gordon, J. J. *et al.* Radiation Pneumonitis and Low Dose Radiation Hypersensitivity. *Proc. World Congr. Med. Phys. Biomed. Eng. Tor. Can. June 7-12 2015*
7. Niemierko, A. & Goitein, M. Modeling of normal tissue response to radiation: the critical volume model. *Int. J. Radiat. Oncol. Biol. Phys.* **25,** 135–145 (1993).
8. Rutkowska, E. S., Syndikus, I., Baker, C. R. & Nahum, A. E. Mechanistic modelling of radiotherapy-induced lung toxicity. *Br. J. Radiol.* **85,** e1242–1248 (2012).
9. Lambin, P., Fertil, B., Malaise, E. P. & Joiner, M. C. Multiphasic survival curves for cells of human tumor cell lines: induced repair or hypersensitive subpopulation? *Radiat. Res.* **138,** S32–36 (1994).
10. Koontz, B. *et al.* Estimation of the alpha/□beta Ratio for Lung Injury Based on Direct Measurements of Radiotherapy (RT)- Induced Reduction in Regional Perfusion in Human Patients. *Int. J. Radiat. Oncol. Biol. Phys.* **63,** S457 (2005).
11. Scheenstra, A. E. H. *et al.* Alpha/Beta Ratio for Lung Tissue as Estimated From Lung Cancer Patients Treated With Stereotactic Body and Conventionally Fractionated Radiation Therapy. *Int. J. Radiat. Oncol. Biol. Phys.* **88,** 224–228 (2014).
12. Borst, G. R. *et al.* Radiation pneumonitis in patients treated for malignant pulmonary lesions with hypofractionated radiation therapy. *Radiother. Oncol.* **91,** 307–313 (2009).
13. Sampath, S., Schultheiss, T. E. & Wong, J. Dose response and factors related to interstitial pneumonitis after bone marrow transplant. *Int. J. Radiat. Oncol. Biol. Phys.* **63,** 876–884 (2005).
14. Reichel, L. & Ye, Q. Simple Square Smoothing Regularization Operators. *Electron. Trans. Numer. Anal.* **33,** 63–83 (2009).
15. Chu, D., Lin, L., Tan, R. C. E. & Wei, Y. Condition numbers and perturbation analysis for the Tikhonov regularization of discrete ill-posed problems. *Numer. Linear Algebra Appl.* **18,** 87–103 (2011).
16. Nelder, J. A. & Wedderburn, R. W. M. Generalized Linear Models. *J. R. Stat. Soc. Ser. Gen.* **135,** 370–384 (1972).
17. Mineault, P. *glmfitqp: Fit GLM with quadratic penalty.* (2014). at <http://www.mathworks.com/matlabcentral/fileexchange/31661-fit-glm-with-quadratic-penalty/content/glmfitqp.m>
18. Guckenberger, M. *et al.* Dose–response relationship for radiation-induced pneumonitis after pulmonary stereotactic body radiotherapy. *Radiother. Oncol.* **97,** 65–70 (2010).

19. Tucker, S. L. *et al.* Impact of toxicity grade and scoring system on the relationship between mean lung dose and risk of radiation pneumonitis in a large cohort of patients with non-small cell lung cancer. *Int. J. Radiat. Oncol. Biol. Phys.* **77,** 691–698 (2010).

20. Metcalfe, P. *et al.* The potential for an enhanced role for MRI in radiation-therapy treatment planning. *Technol. Cancer Res. Treat.* **12,** 429–446 (2013).

21. Ma, J. *et al.* Regional lung density changes after radiation therapy for tumors in and around thorax. *Int. J. Radiat. Oncol. Biol. Phys.* **76,** 116–122 (2010).

22. Zhang, J. *et al.* Radiation-induced reductions in regional lung perfusion: 0.1-12 year data from a prospective clinical study. *Int. J. Radiat. Oncol. Biol. Phys.* **76,** 425–432 (2010).

23. Liao, Z., Travis, E. L. & Komaki, R. in *Chapter 44, Principles and Practice of Lung Cancer: The Official Reference Text of the International Association for the Study of Lung Cancer, H.I. Pass et al, Lippincott Williams & Wilkins, 2012.*

24. Siva, S. M. *et al.* Abscopal Effects after Conventional and Stereotactic Lung Irradiation of Non-Small-Cell Lung Cancer. *J. Thorac. Oncol.* **8,** (2013).

25. Song, S. H., Oh, Y. J., Ha, C. W. & Chae, S.-M. Bilateral Diffuse Radiation Pneumonitis Caused by Unilateral Thoracic Irradiation: A Case Report. *J Lung Cancer* **11,** 97–101 (2012).

26. Gómez-Millán, J. *et al.* The importance of bystander effects in radiation therapy in melanoma skin-cancer cells and umbilical-cord stromal stem cells. *Radiother. Oncol.* **102,** 450–458 (2012).

27. Formenti, S. C. & Demaria, S. Combining radiotherapy and cancer immunotherapy: a paradigm shift. *J. Natl. Cancer Inst.* **105,** 256–265 (2013).

28. Ramsay, J. & Silverman, B. W. *Functional Data Analysis.* (Springer, 2005).

**Corresponding author:**
Author:       J.J. Gordon
Institute:   Henry Ford Dept. Radiation Oncology
Street:       2799 W Grand Boulevard
City:          Detroit MI 48202
Country:    U.S.A.
Email:        jjgw@jjgordon.com

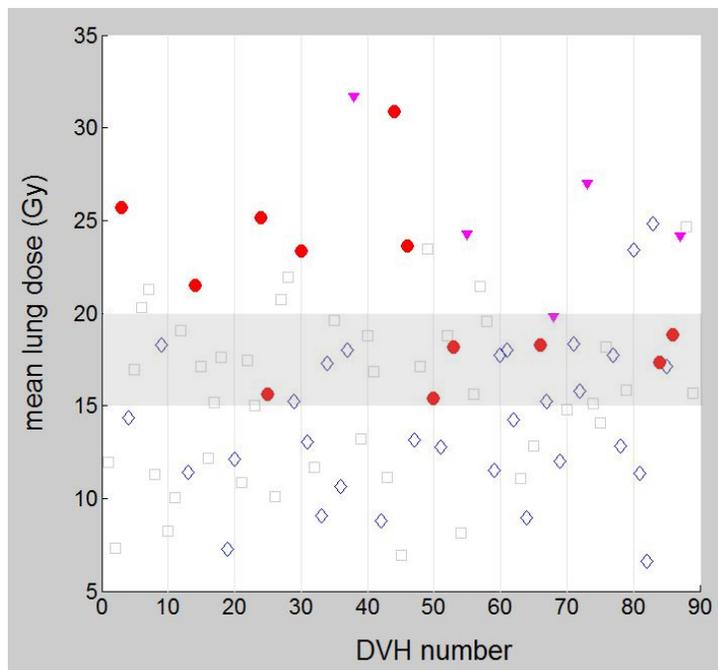

**Figure 1:** Plot of mean lung dose (MLD) versus DVH number for the University of Michigan dataset. Actual radiation pneumonitis (RP) incidence, derived from the clinical trial, is indicated by symbol. Pink solid triangles are grade 3 DVHs. Red solid circles are grade 2 DVHs. Blue hollow diamonds are grade 1 DVHs. Gray hollow squares are grade 0 DVHs. Note that these toxicity gradings are ignored in the simulated trials. In the simulations, RP incidence is generated using equations (2-4). The gray band indicates the subset of DVHs (i.e., those having 15 Gy ≤ MLD ≤ 20 Gy) which was used for some simulations.

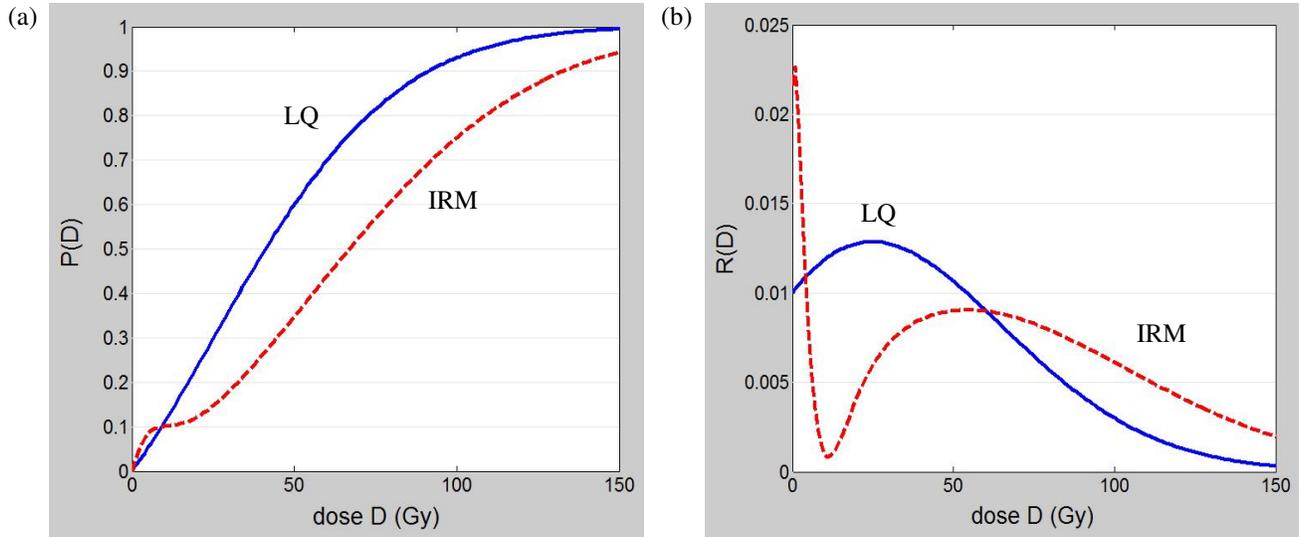

**Figure 2:** **(a)** Dose-damage profiles P(D) corresponding to the linear quadratic (LQ) model (blue solid line) and Induced Repair Model (IRM) (red dashed line). **(b)** Rate profiles R(D) corresponding to the LQ and IRM models. The rate profile is the derivative of the damage profile: R(D) = P′(D).

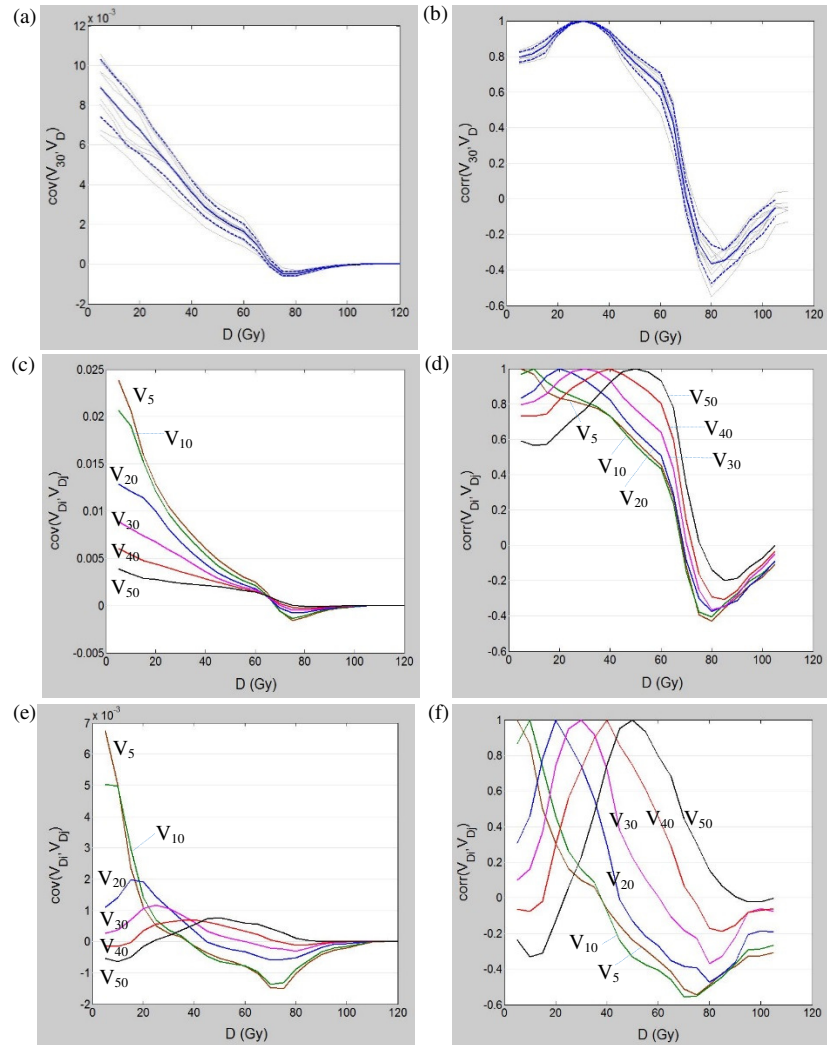

**Figure 3:** Plots of covariance and correlation (CC) between dose metrics VD. **(a-b)** CC between V30 and VD, where D = 5, 10, ... , 100 Gy (x-axis). The gray lines are the CC obtained for 10 simulated clinical trials, each containing 100 DVHs. The central solid blue line is the average. The outer dashed blue lines represent the average ± one standard deviation. **(c-d)** Average CC between $VD_i$ and $VD_j$, where simulated DVHs are generated from the full UM DVH set. **(c-d)** Average CC between $VD_i$ and $VD_j$, where simulated DVHs are restricted to the MLD range 15 Gy ≤ MLD ≤ 20 Gy.

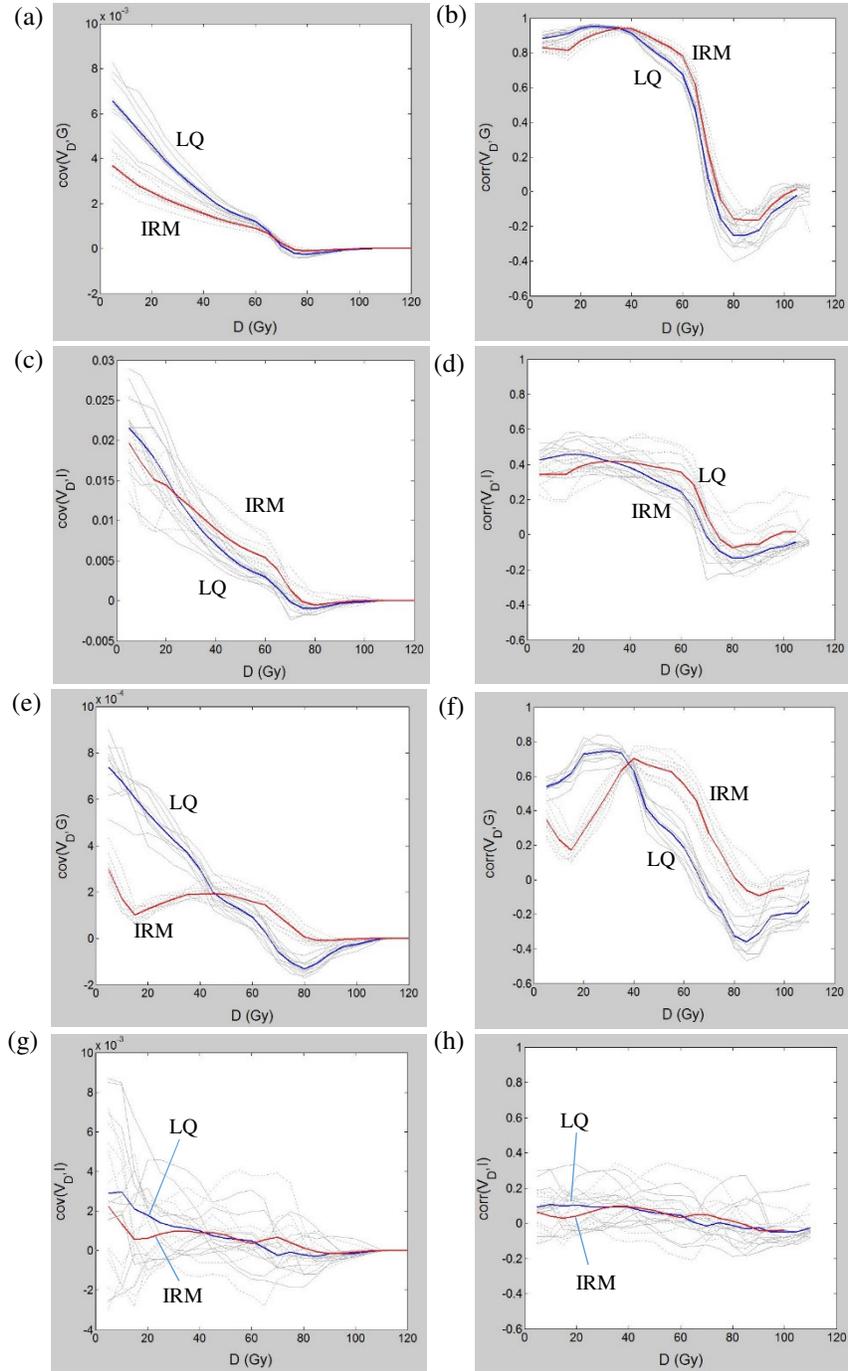

**Figure 4:** Plots of covariance and correlation (CC) between G, I and dose metrics VD. Gray lines are for individual trials. Blue and red lines are averages over the trials. The solid gray and blue lines are for the LQ dose-response model. The dotted gray and red solid lines are for the IRM model. **(a-d)** CC for the case where simulated DVHs are generated from the full UM DVH set. **(e-h)** CC for the case where simulated DVHs are restricted to the MLD range 15 Gy ≤ MLD ≤ 20 Gy.

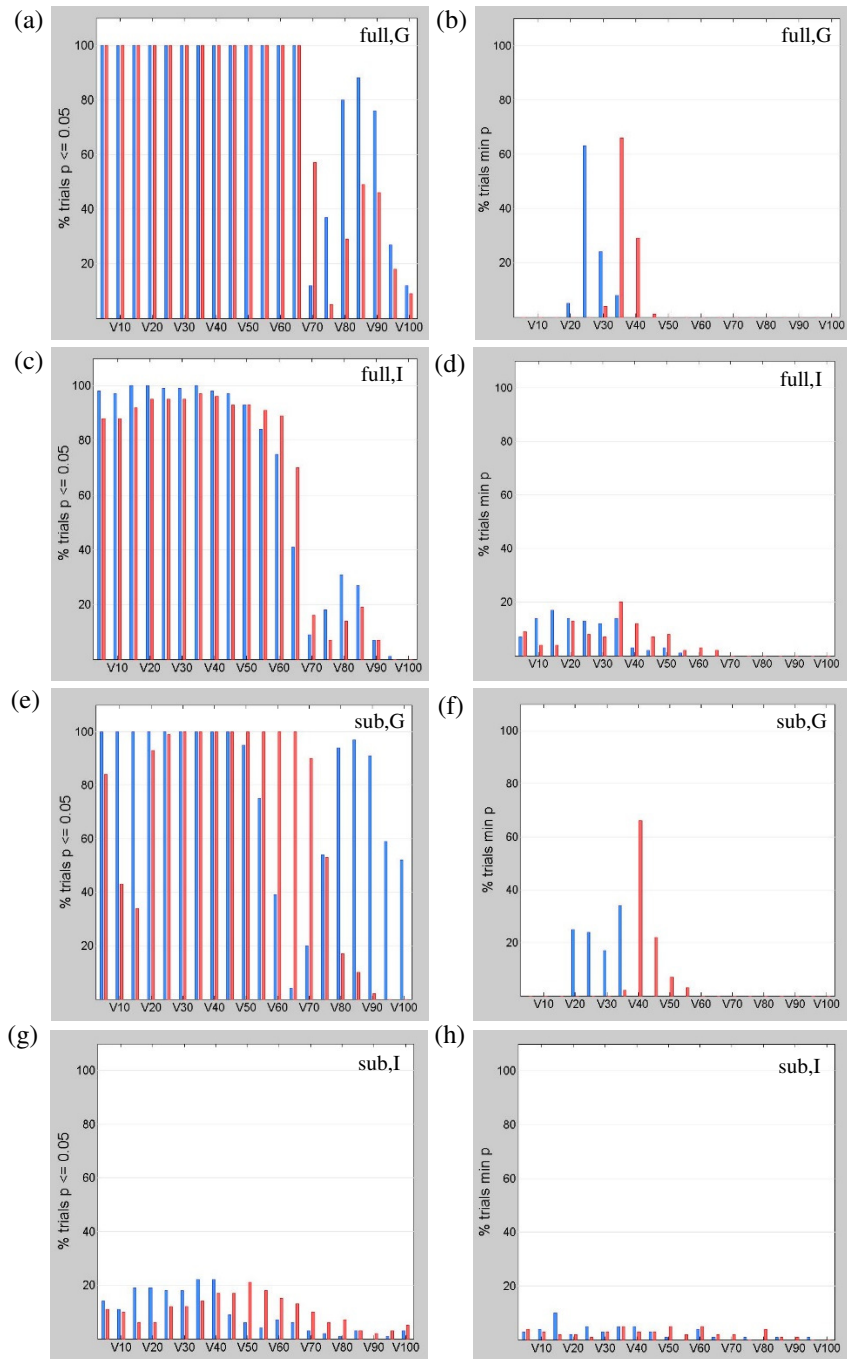

**Figure 5: (a,c,e,g)** Plots of the percentage of trials in which VD has a statistically significant (p < 0.05) correlation with G or I. **(b,d,f,h)** Plots of the percentage of trials in which VD has the most significant (smallest p-value) correlation with G or I. Panels a-d are for the case where simulated DVHs are generated from the full UM DVH set (labelled "full"). Panels e-h are for the case where simulated DVHs are restricted to the MLD range 15 Gy ≤ MLD ≤ 20 Gy (labelled "sub"). Blue bars are for the LQ dose-response model. Red bars are for the IRM dose-response model.

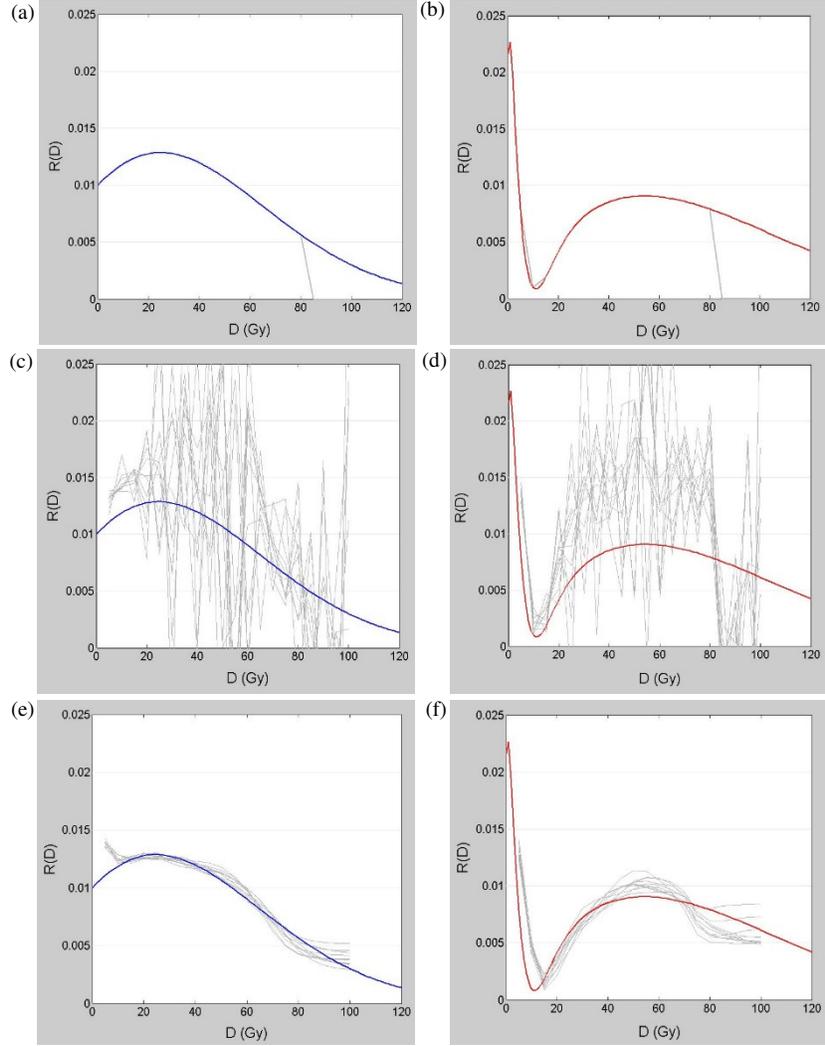

**Figure 6:** Estimates of R(D) obtained using matrix methods. Panels a,c,e show results for the LQ profile (blue line). Panels b,d,f show results for the IRM profile (red line). Gray lines are the 10 estimates of R(D) obtained from 10 trials x 100 DVHs. Results are for the full DVH dataset. Results for the restricted DVH dataset having 15 Gy ≤ MLD ≤ 20 Gy are visually similar. **(a-b)** Solutions $\tilde{r} = M^{-1}\,\tilde{c}$ for the case where there is zero noise (i.e., R(D) is replaced by R$_{ZN}$(D) in the damage calculations). **(c-d)** Solutions $\tilde{r} = M^{-1}\,\tilde{c}$ for the case where small amounts of noise are added to G. **(e-f)** Solutions of the Tikhonov regularized equation (15).

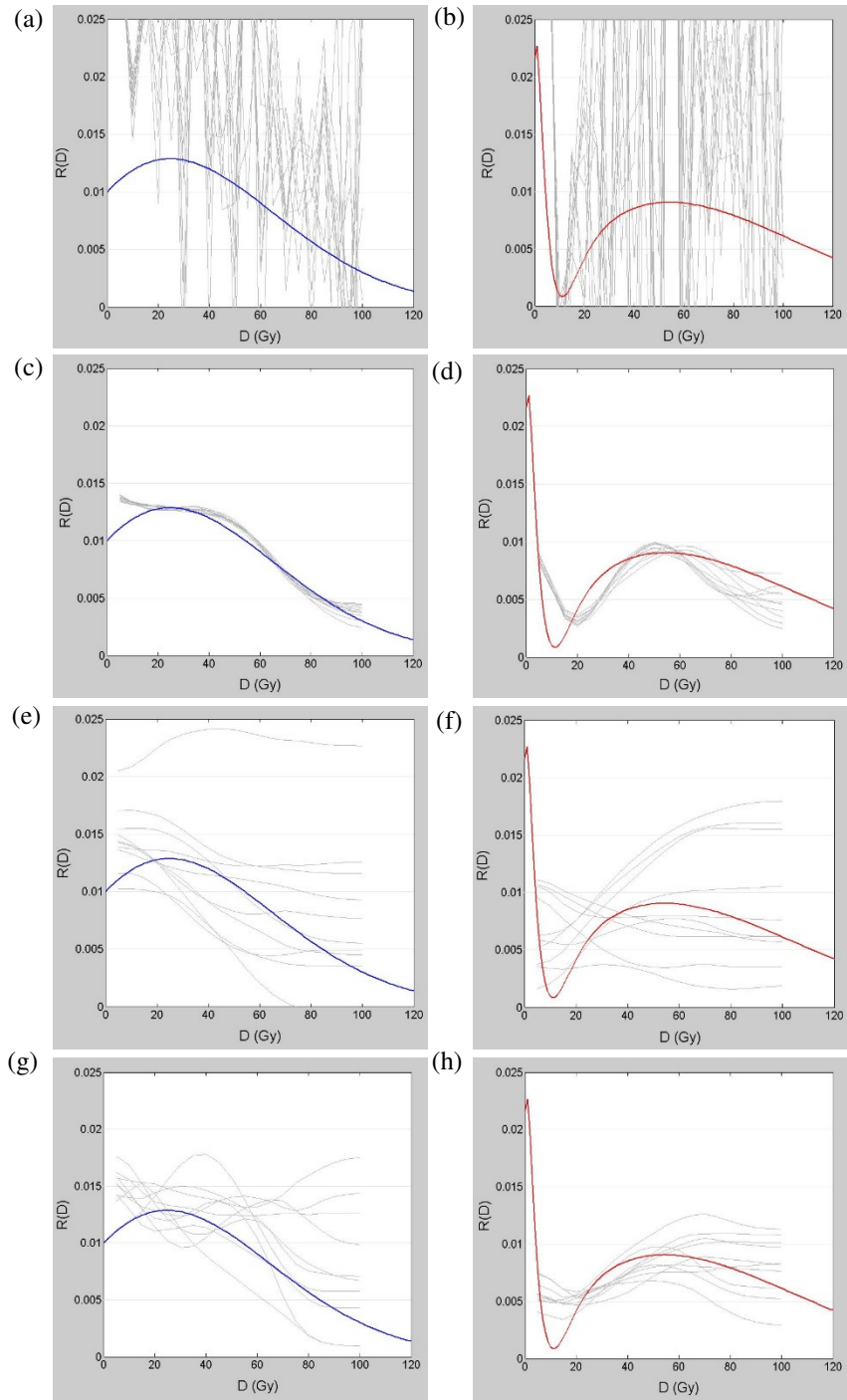

**Figure 7:** Regression solutions for simulated data. **(a-b)** G vs VD using Matlab's *glmfit()* function. **(c-d)** G vs VD using Mineault's *glmfitqp()* function. **(e-f)** I vs VD using Mineault's *glmfitqp()* function, 10 trials x 100 DVHs. **(g-h)** I vs VD using Mineault's *glmfitqp()* function, 10 trials x 1000 DVHs.

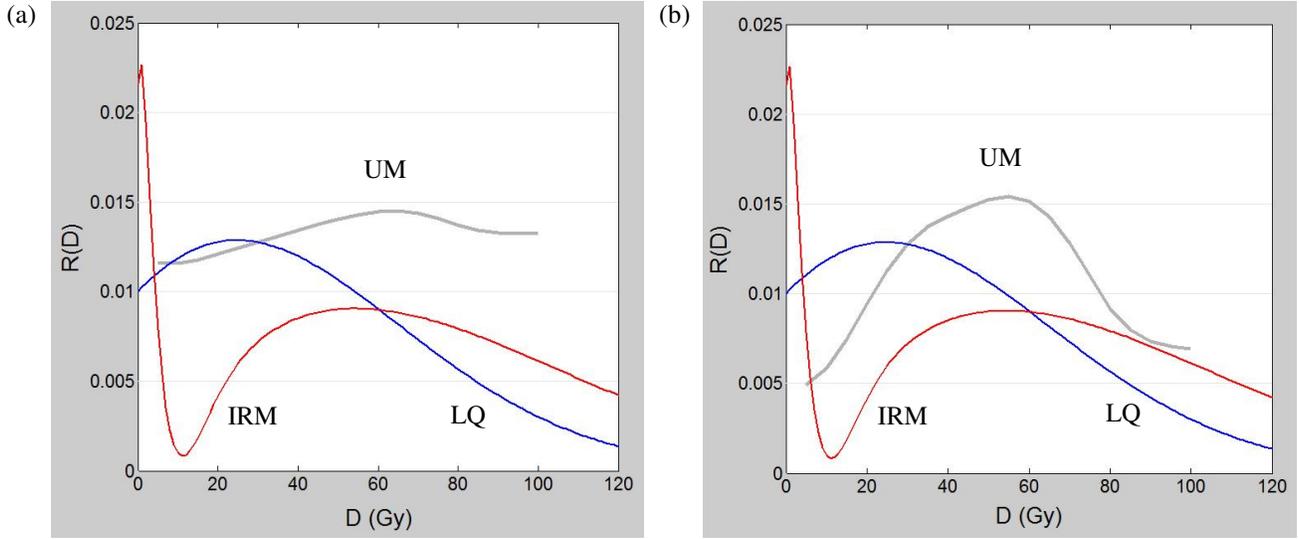

**Figure 8:** Regression solutions for University of Michigan clinical trial data. **(a)** R(D) extracted using full DVH dataset (gray line), compared with LQ model (blue solid line) and IRM model (red line). **(b)** R(D) extracted using restricted DVH dataset having 15 Gy ≤ MLD ≤ 20 Gy.